\documentstyle{l-aa}
\begin{document}

\thesaurus{08
              (08.13.1;
               08.14.1)
               13
               (13.07.1;
               13.25.5)}

\title{On the persistent X-ray emission from the soft $\gamma$-ray
repeaters
}
\author{V.V.~Usov}
\offprints{V.V. Usov}
\institute{
Department of Condensed Matter Physics, Weizmann Institute,
Rehovot 76100, Israel}
\date{Received  $\,\,\,\,\,\,\,\,\,\,\,\,\,\,\,\,\,\,\,\,\,$ / Accepted   }
\maketitle

\begin{abstract}
It is suggested that the persistent X-ray emission from the soft
$\gamma$-ray repeaters is the thermal radiation of neutron stars
which is enhanced by a factor of 10 or more due to
the effect of a very strong magnetic field on the thermal structure of
the neutron star envelope. For the thermal luminosity to be consistent
with the persistent X-ray luminosity, the field strength
at the neutron star surface has to be of the order of
$10^{15}$ G. If it is confirmed that the soft $\gamma$-ray repeaters
are neutron stars with negligible accretion, then
the presence of such a strong magnetic field is inevitable.

\end{abstract}

\section{Introduction}

The soft $\gamma$-ray repeaters (SGRs) are a small, enigmatic class of
$\gamma$-ray transient sources (Norris et al. \cite{Norris1991}). There
are three known SGRs: 0526-66, 1806-20 and 1900+14. All these repeaters
have been associated with young supernova remnants. The first, SGR
0525-66, which is the source of the well-known 1979 March 5 burst, is
associated with the N49 supernova remnant (SNR) in the Large Magellanic
Cloud (Cline et al. \cite{Cline1982}). The second, SGR 1806-20, is
located toward the Galactic Center, and is associated with SNR
G10.0-0.3 (Kulkarni et al. \cite{Kulkarni1994}). The age of both N49 and
G10.0-0.3 is about $5\times 10^3$ yr. The third, SGR 1900+14, is
associated with SNR G42.8+0.6, and its age is $\sim 10^4$ yr (Vasisht
et al.\cite{Vasisht1994}). The positions of SGRs are offset from the
centers of their SNRs. This implies a very high velocity, up to $\sim
1200$ km s$^{-1}$ or even more, for SGRs.  Accepting these SGR - SNR
associations, the corresponding burst peak luminosities were estimated.
If the burst radiation is more or less unbeamed, these luminosities are
a few orders higher than the standard Eddington value for a star with
the mass of $M_\odot$. For example, SGR 1806-20 has produced events
that are $\sim 10^4$ times the Eddington luminosity (Fenimore, Evans \&
Ulmer \cite{Fenimore1994}). The durations of these events range
between 0.1 s and 200 s.
In addition to short bursts of both hard X-rays and soft
$\gamma$-rays, the persistent X-ray emission was detected from SGRs as
well. (Rothschild, Kulkarni \& Lingenfelter \cite{Rothschild1994};
Murakami et al.\cite{Murakami1994}; Visisht et al. \cite{Vasisht1994};
Hurley et al. \cite{Hurley1996}). The luminosity of the persistent
X-ray sources is $\sim 7\times 10^{35}$ erg s$^{-1}$ for SGR 0525-66,
$\sim 3\times 10^{35}$ erg s$^{-1}$ for SGR 1806-20, and $\sim 10^{35}$
erg s$^{-1}$ for SGR 1900+14. In combination, these results support the
preposition that SGRs can be firmly identified as neutron stars. High
velocities of SGRs imply that the neutron stars are single. It was
argued (Usov \cite{Usov1984}; Duncan \& Thompson \cite{DT1992};
Paczy\'nski \cite{Pac1992}; Thompson \& Duncan \cite{TD93},
\cite{TD95}; Podsiadlowski, Rees \& Ruderman \cite{PRR95})
that not very young, single neutron stars may be a source of
strong bursts of high-frequency, hard X-ray and $\gamma$-ray emission
only if their magnetic fields are very high, $B\sim 10^{14}$ G or more
(maybe, up to $\sim 10^{16}$ G).

Calculations for the cooling of neutron stars predict
(e.g., Nomoto \& Tsuruta \cite{NT1987}) that after $(0.5-1)\times
10^4$ yr the bolometric luminosities will be at least an order
less than the X-ray luminosities of the persistent X-ray sources
which are identified with SGRs. At first sight, a source of heating,
such as accretion of gas, is necessary to account for the higher
thermal luminosities (e.g., Rothschild et al. \cite{Rothschild1994}).
But this is not always a prerequisite. In this letter, it is argued
that a very strong magnetic field, $\sim 10^{14}-10^{16}$ G,
can influence the physical conditions in the surface layers of SGRs so
that the thermal luminosities of the neutron stars may increase up to
observed luminosities of the persistent X-ray sources.

\section{Surface structure}

The structure of matter in the surface layers of neutron stars with
the surface field $B\gg\alpha^2B_{\rm cr}\simeq 2.35\times
10^9$ G is largely determined by the magnetic field,
where $\alpha=e^2/\hbar c=1/137$ is the the fine structure
constant and $B_{\rm cr}=m^2c^2/e\hbar\simeq 4.4\times 10^{13}$ G
(Ruderman \cite{Rud1971};
Flowers et al. \cite{F1977}; Fushiki, Gudmundsson and Pethick
\cite{FGP1989}; Abrahams \&
Shapiro \cite{AS1991}; R\"ognvaldsson et al. \cite{RF1993}). It has
been suggested that the surface of neutron stars with such a strong
magnetic field consists of magnetic metal in which atoms form
chains aligned along the field lines (Chen, Ruderman \& Sutherland
\cite{CRS1974}).

The strength of a magnetic field on an atom of atomic number $Z$ is
characterized by the following dimensionless parameter
(Ruderman \cite{Rud1971}):

\begin{equation}
\eta = \left({B\over 2\alpha^2B_{\rm cr}Z^3}\right)^{1/2}
=\left({B\over 4.7\times 10^9Z^3\,{\rm G}}\right)^{1/2}\,.
\label{eta}
\end{equation}

In the case of very strong magnetic fields, $\eta\gg 1$, as it is
applied to SGRs, some simplifying assumptions
can be made about the electronic structure of atoms and condensed
matter. In this case, the {\it adiabatic approximation} is valid:
the magnetic field completely determines the transverse electron
motion. This motion is quantized into Landau levels, and only the
lowest Landau level is populated (this assumes temperatures such
that $kT\ll \Delta \varepsilon_{01}$, where $\Delta
\varepsilon_{01}=mc^2[\sqrt{1+(2B/B_{\rm cr})} -1]$ is the energy
difference between the first excited Landau level and the lowest one).

The surface of a neutron star with a very strong magnetic field,
$\eta\gg
1$, is probably a magnetic metal, provided that the surface
temperature
is smaller than the melting temperature (Usov \& Melrose
\cite{UM1995} and references therein). The density of
the magnetic metal phase is insensitive to the atomic geometry. At
the neutron star surface, this density is (Flowers et al.
\cite{F1977})

\begin{equation}
\rho_{_{\rm S}}(B)\simeq 4\times 10^3
\left({A\over 56}\right)\left({Z\over 26}\right)^{-3/5}
\left({B\over 10^{12}\,{\rm G}}\right)^{6/5}{\rm g\,cm}^{-3},
\label{rho}
\end{equation}

\noindent where $A$ is the mass number of atoms.

The composition of the
neutron star surface is uncertain and depends on the star's history.
If the neutron star has never accreted matter
onto its surface, it is natural to expect that the neutron star
surface consists almost entirely of $^{56}$Fe, $A=56$ and $Z=26$
(e.g.,
Usov \& Melrose \cite{UM1995}). In this case, the surface field is
very strong, $\eta\gg 1$, when its strength is much higher than
$8\times
10^{13}$ G.

For intermediate magnetic fields, $Z^{-3/2}\ll \eta \ll 1$,
the qualitative result of Cheng et al. (\cite{CRS1974}),
that chains are energetically favoured over
individual atoms, is questionable. The calculations of Neuhause,
Koonin \& Langanke (\cite{NKL1987}) indicate that at $B\la 10^{13}$ G
free atoms of $^{56}$Fe
are preferred over chains, and iron does not form a
magnetic metal (however, see Abrahams \& Shapiro \cite{AS1991}). In
this
case, the stellar surface is usually identified with the base of the
photosphere which is at an optical depth $\tau ={2\over 3}$. The
photon spectrum mainly forms at this depth.
Strong magnetic fields significantly influence the transport
properties
of neutron-star atmospheres (e.g., Hernquist \cite{H1985}). For
example,
such a field can strongly reduce the opacity. As a result, at
$B\gg \alpha^2B_{\rm cr}$ the surface density found by the radiative
boundary condition, $\tau ={2\over 3}$,
increases many orders because of the effect of a strong magnetic field
(e.g., Van Riper \cite{V1988}). Hence, the density $\rho_{_{\rm
S}}(B)$
at the surface of neutron stars with strong magnetic
fields is much higher than the surface density $\rho_{_{\rm
S}}(B=0)\sim
0.1-0.01$ g cm$^{-3}$ in the zero-field case regardless of
formation of the magnetic metal phase.

All mentioned studies of iron matter in the field $B\gg
\alpha^2B_{\rm cr}$ were based on nonrelativistic quantum mechanics.
For $B\ga B_{\rm cr}$, the transverse motion of electrons becomes
relativistic. However, as noted earlier, in very strong magnetic
fields,
$\eta\gg 1$,
the motion of electrons perpendicular to the field is frozen into the
lowest Landau level, which is rather insensitive to relativistic
corrections. These corrections to the matter parameters are
small at $B>B_{\rm cr}$ as long as the electron motion remains
nonrelativistic
along the field direction (Angelie \& Deutsch \cite{AD1978}; Lai \&
Salpeter \cite{LS1995}). In the instance of condensed iron matter,
the electron motion along the field may be roughly considered
as nonrelativistic at least up to the field strengths of
a few $\times 10^{15}$ G (Glasser \& Kaplan \cite{GK1975}).

\section{Thermal luminosity}

If neutron stars are not too young (age $t\ga 10^2$ yr),
for consideration of the thermal structure and
photon radiation, it is convenient to divide the stellar interior
into two regions: the isothermal core with density $\rho >\rho_e$ and
the outer envelope with $\rho<\rho_e$, where $\rho_e\sim 10^{10}$ g
cm$^{-3}$ (Gudmundsson, Pethick \& Epstein \cite{GPE1983}). At
$t\simeq (0.5-1)\times 10^4$ yr, for a typical neutron star
[$M\simeq 1.4M_\odot$, $R\simeq 1.6\times 10^6$ cm, and the bulk of
neutrons in the core is superfluid] with $B=0$ at the surface
{\it standard} cooling calculations
yield that the temperature decreases by a factor of $2\times 10^2$
in the envelope, from $T_c\simeq 4\times 10^8$ K at the inner
boundary of the envelope to $T_{_{\rm S}}\simeq 2\times 10^6$ K at
the neutron star surface (Nomoto \& Tsuruta \cite{NT1987}).
(By {\it standard} cooling we mean cooling
without such exotic particles as charged pion condensates and quarks
which might prove to be very fast cooling agents.) In this case the
expected photon luminosity is $L_{\rm ph}(B=0)\simeq
10^{34}$ erg s$^{-1}$ within a factor 2 or so.

At $B\gg \alpha^2B_{\rm cr}$ the surface density
increases in comparison with $\rho_{_{\rm S}}(B=0)$.
Besides, the effect of
quantization of electron orbits increases the longitudinal thermal
conductivity of degenerate electrons. Both these effects may result
in an increase of the surface temperature (and the photon luminosity)
when the neutron star is not too old ($t<3\times 10^5$ yr) and cools
mainly
via neutrino emission from the stellar interior (e.g.,  Van Riper
\cite{V1988}, \cite{V1991}; Shibanov \& Yakovlev \cite{SY1996}).

The effect of a strong magnetic field on neutron-star cooling has been
considered by a number of authors (e.g., Tsuruta \cite{T1979};
Van Riper \& Lamb \cite{VL1981}; Yakovlev \& Urpin \cite{YU1981};
Nomoto \& Tsuruta \cite{NT1987}; Schaaf \cite{S1990}; Page
\cite{P1995};
Shibanov \& Yakovlev \cite{SY1996}). According to early calculations
of Tsuruta (\cite{T1979}), Van Riper \& Lamb (\cite{VL1981}), Yakovlev
\& Urpin (\cite{YU1981}) and others, the magnetic field effect can be
quite significant even if the field strength is not very high, $B\la
B_{\rm cr}$. For example, at the core temperature $T_c=10^8$ K the
predicted enhancements in the flux of radiation (which are the same as
the luminosity enhancement) are factors $\sim 10$ for
$B\simeq 0.1 B_{\rm cr}$ and $\sim 10^2$ for $B\simeq B_{\rm cr}$.
However, the effect of strong magnetic fields has been overestimated
in these calculations (Hernquist \cite{H1985}).
This is because the thermal conductivities
which have been used are not accurate, and the tensor nature of the
photon
transport has been ignored. Besides, all early calculations have been
performed under the simplified assumption that the magnetic field is
radial everywhere over the stellar surface. Using more realistic
heat-transport coefficients it was shown that for a fixed core
temperature and a purely radial field of $B\sim 0.1B_{\rm cr}$
the effects of strong
magnetic fields increase the heat flux (relative to the zero-field
case) by a factor $\la 3$. In turn, at $B\ga 10^{12}$ G the
electron thermal conductivity across the field is strongly suppressed,
and for a realistic field geometry the mean heat flux
is reduced by a factor $2-3$. Thus, it was concluded that
at the neutrino cooling stage of neutron stars with
intermediate magnetic fields, $B\sim 0.1B_{\rm cr}$,
the enhancement in the heat flux due to the
magnetic field effects
will be approximately cancelled by the suppression of the heat
flux due to the overall field geometry (Hernquist \cite{H1985}). This
is reasonably congruent with the {\it standard} cooling calculations
of Shibanov \& Yakovlev (\cite{SY1996}) for a typical neutron star
with a dipole magnetic field.
At $t \simeq (0.5-1)\times 10^4$ yr, the enhancement in the
photon luminosity (relative to the zero-field case)
is a factor $\sim 2$ for $B=10^{13.5}$ G,
$L_{\rm ph}(B=10^{13.5}\,{\rm G})\simeq 2L_{\rm ph}(B=0)$
(Shibanov \& Yakovlev \cite{SY1996}). When the strength of the surface
magnetic field changes from $B=10^{13.5}$ G to $B=10^{14}$ G, the
surface temperature changes by a factor $\sim 1.2$
(Van Riper \cite{V1988}). Hence, at $t \simeq (0.5-1)\times 10^4$ yr
and
$B=10^{14}$ G the expected enhancement in the photon luminosity is a
factor $\sim 4$, $L_{\rm ph}(B=10^{14}\,{\rm G})\simeq 4L_{\rm
ph}(B=0)$.
The enhancement in the photon luminosity
is more or less the same irrespective of whether the formation of
magnetic metal is at the neutron star surface or not (Van Riper
\cite{V1988}).

In all available numerical calculations the thermal structure
of magnetized neutron stars were studied for
$B\leq 10^{14}$ G. However, the input physics for such calculations
does not qualitatively change up to the field strengths of $B\simeq$ a
few $\times 10^{15}$ G. Therefore,
the tendency of the photon luminosity to increase with increase
of $B$ has to be held at least up to
$B\simeq$ a few $\times 10^{15}$ G.
The maximum photon luminosity, $L^{\rm max}_{\rm ph}(t)$, of a
magnetized
neutron star with the age $t$ may be estimated in the following way.
As noted earlier, the photon luminosity can be enhanced
by the effect of strong magnetic fields only at the
neutrino cooling stage. At this stage, for a typical
neutron star with the {\it standard} neutrino energy losses,
the neutrino luminosity is (Nomoto \& Tsuruta \cite{NT1987})

\begin{equation}
L_\nu(t)\simeq 10^{36}\left({t\over 5\times 10^3\,{\rm
yr}}\right)^{-4/3}
\,\,\,{\rm erg\,s}^{-1}\,.
\label{Lnu}
\end{equation}

\noindent The photon luminosity may be enhanced
by the field effect up to the neutrino luminosity, i.e.,
$L^{\rm max}_{\rm ph}(t)\simeq L_\nu(t)$. This may be done
because the enhancement in the photon luminosity may be, in principle,
as high as a factor $\sim (T_c/T_{_{\rm S}})^4\sim 10^8$.
For all SGRs, $L^{\rm max}_{\rm ph}(t)$
is more than the observed luminosities of the persistent X-ray
sources.
Let us estimate the minimum photon luminosity of a magnetized
neutron star at the neutrino cooling stage.

>From the calculations of Van Riper (\cite{V1988}) it follows that the
temperature change, $T_c/T_{_{\rm S}}$, through the envelope is
strongly reduced at $T_{_{\rm S}}<T_0$, where

\begin{equation}
T_0\simeq 2.2\times 10^5\left({B\over 10^{12}\,{\rm
G}}\right)^{1/3}\,\,
\,{\rm K}\,.
\label{T0}
\end{equation}

At the neutrino cooling stage,
$T_{_{\rm S}}$ cannot be essentially smaller than $T_0$. Indeed, at
this stage the core temperature does not depend on $T_{_{\rm S}}$
(e.g., Nomoto \& Tsuruta \cite{NT1987}). If, in the process of the
neutron star cooling, the surface temperature $T_{_{\rm S}}$ becomes
smaller than $T_0$, the ratio $T_c/T_{_{\rm S}}$ has to be strongly
reduced. However, it is possible only if $T_c$ is strongly reduced
too.
In turn, such a reduction of $T_c$ is possible only at the photon
cooling stage. Hence, for a typical neutron star with the field $B$ at
the surface the photon luminosity at the neutrino cooling stage
cannot be essentially smaller than

\begin{equation}
L_{\rm ph}^{\rm min}\simeq
4\pi R^2\sigma T_0^4\simeq 3\times 10^{35}\left({B
\over 4\times 10^{15}\,{\rm G}}\right)^{4/3}\,\,{\rm erg\,s}^{-1},
\label{L0}
\end{equation}

\noindent where $\sigma$ is the Stefan-Boltzmann constant. Certainly,
this estimate may be used only if $L_{\rm ph}^{\rm min}<L_\nu$.

\section{Conclusions and discussion}

It is suggested in this letter that the persistent X-ray emission from
SGRs is the thermal radiation of neutron stars which is enhanced by
the
effects of a very strong magnetic field on the thermal structure of
the neutron star envelope. Since for a nonmagnetic neutron star at
$t \simeq (0.5-1)\times 10^4$ yr we have $L_{\rm ph}(B=0)\la 2\times
10^{34}$ erg s$^{-1}$, the field $B=10^{14}$ G,
for which the expected enhancement in the photon luminosity is a
factor
$\sim 4$, is a firm lower limit on the field strength at the surface
of SGRs. The enhanced luminosity in thermal X-rays
may be as high as the neutrino
luminosity. At $t \simeq (0.5-1)\times 10^4$ yr, for a typical neutron
star the neutrino luminosity obtained from the {\it standard} cooling
theory is more than the persistent X-ray luminosities of SGRs. If our
extrapolation of a strong decrease of the temperature change,
$T_c/T_{_{\rm S}}$, through the envelope at $T_{_{\rm S}}<T_0$
into the region of very strong magnetic fields, $\sim 10^{14}-10^{16}$
G,
is correct, we have the following upper limit on $B$:
$B \la 7\times 10^{15}$ G (see,
equation (\ref{L0})). Hence, the strength of the field at the surface
of SGRs is somewhere between a few $\times 10^{14}$ G and a few
$\times 10^{15}$ G. This is consistent with the estimates of Thompson
\& Duncan (\cite{TD95}) for the magnetic field at the surface of SGRs.

Some effects are ignored in our
consideration. One of them is the effect of strong magnetic fields on
the energy losses via neutrino emission from the core of the neutron
stars,
which may be essential at $B \ga 10^{15}$ G (Cheng, Schramm \&
Truran \cite{CST93}). Another is the magnetic field decay. The field
strength required to power the persistent X-ray emission
for $\sim 10^4$ yr is $\sim 10^{15}$ G or more (Thompson
\& Duncan \cite{TD95}). If accretion onto the neutron stars
is negligible, both the thermal energy and the
magnetic field energy could be a source of energy for the
persistent X-ray emission. In both these cases the field strength has
to be $B\ga 10^{15}$ G.

The presence of {\it exotic} particles such as a pion condensate in
the
core of neutron stars can result in extremely rapid cooling
(e.g., Richardson \cite{R1982}). For such {\it nonstandard} rapid
cooling, at $t \simeq (0.5-1)\times 10^4$ yr the expected neutrino
luminosity is many orders smaller than $L_\nu$ given by equation
(\ref{Lnu}), that excludes the thermal radiation from the neutron star
surface as a source of the persistent X-ray emission from SGRs.

\begin{acknowledgements}
I thank K. Hurley for sending his paper which stimulated this work.
I also thank M. Milgrom for helpful conversations, and M. Waller and
O. Usov for careful reading of the manuscript.
\end{acknowledgements}

\end{document}